\documentclass[12pt]{article}
\include{epsf}

\newcommand{\eqn}[1]{(\ref{#1})}
\def\appendix#1{\addtocounter{section}{1}\setcounter{equation}{0}
\renewcommand{\thesection}{\Alph{section}}
\section*{
\thesection\protect\indent \parbox[t]{11.715cm} {#1}}
\addcontentsline{toc}{section}{Appendix\thesection\ \ \ #1} }
\newcommand{\real}{{\bb R}} 
\newcommand{\reals}{{\bbs R}} 
\newcommand{\id}{{1\!\!1}} 

 \font\mybb=msbm10 at 12pt
\def\bb#1{\hbox{\mybb#1}}
\font\mybbs=msbm10 at 9pt
\def\bbs#1{\hbox{\mybbs#1}}

\def\nn{\nonumber}

\def\be{\begin{equation}}
\def\ee{\end{equation}}
\def\beqa{\begin{eqnarray}}
\def\eeqa{\end{eqnarray}}

\newcommand{\half}{{\textstyle{1\over 2}}}

\newcommand{\del}{\partial}

\begin{document}
\begin{titlepage}
\begin{flushright}

\baselineskip=12pt
DSF--17--2004\\ hep--th/yymmnn\\
\hfill{ }\\
October 2004\\
revised April 2005
\end{flushright}

\begin{center}

\baselineskip=24pt

{\Large\bf Noncommutative differential calculus for Moyal
subalgebras }

\baselineskip=14pt

\vspace{1cm}

{{\bf G.~Marmo, P.~Vitale and A.~Zampini}}
\\[6mm]
 {\it Dipartimento di Scienze Fisiche, Universit\`{a} di
Napoli {\sl Federico II}\\ and {\it INFN, Sezione di Napoli}\\
Monte S.~Angelo, Via Cintia, 80126 Napoli, Italy}\\ {\tt
giuseppe.marmo, patrizia.vitale, alessandro.zampini@na.infn.it}
\\[10mm]

\end{center}

\vskip 2 cm

\begin{abstract}
We build a differential calculus for subalgebras of the Moyal
algebra on $R^4$ starting from a redundant differential calculus
on the Moyal algebra, which is suitable for reduction. In some
cases we find a frame of 1-forms which allows to realize the
complex of forms as a tensor product of the noncommutative
subalgebras with the external algebra  $\Lambda^\ast$.\\
{\tt MSC: 46L87;\\ Keywords: Noncommutative differential
calculus.}
\end{abstract}

\end{titlepage}

\section{Introduction}
In this paper we address the problem of building a differential
calculus on a wide class of noncommutative algebras introduced in
\cite{GLMV02}.  Those are inequivalent infinite-dimensional
$\ast$-algebras  in one to one correspondence  with subalgebras of
the Moyal algebra on  $R^4$, which all share the same commutative
limit, namely the algebra of functions on $R^3$. Following some
ideas of Segal contained in \cite{segal}, where he defines  a {\it
Quantized Differential Calculus} for the algebra of operators of
Quantum Mechanics, we build a differential calculus based on the
existence of a sufficient number of derivations. The algebras we
are interested in are subalgebras of a bigger one, therefore, in
our approach, an important point is how to infer a differential
calculus for subalgebras from a given differential calculus on the
big algebra. The problem is nontrivial in the noncommutative case,
and of interest also in more general situations where we have
morphisms between two algebras ${\mathcal A}, {\mathcal B}$, which
could be, for example, the noncommutative analogues of the source
and target space of field theories. In the commutative case, given
M, N, a pair of  differentiable manifolds with some $\phi:
M\rightarrow N$, we know that the exterior derivative on the two
spaces is connected by a pull-back \be \phi^*(d_N f ) = d_M
\phi^*( f)\label{pullbackd}\ee where $f\in {\mathcal F}(N)$ while
$\phi^*( f)\in {\mathcal F}(M)$. But, if the commutative algebras
of functions ${\mathcal F}(M), {\mathcal F}(N)$ are replaced by
the noncommutative algebras ${\mathcal A}, {\mathcal B}$ with some
$\psi: {\mathcal B}\rightarrow {\mathcal A}$ the relation between
the differential calculi on the two algebras is not obvious a
priori. Can we use the differential calculus on ${\mathcal A}$ to
define a differential calculus on ${\mathcal B}$ as in
Eq.~\eqn{pullbackd}? As we shall see, this is in general not
possible, essentially because derivations in the noncommutative
case are not a ${\mathcal A}$-module, namely we cannot multiply
them by elements of ${\mathcal A}$ so that they remain
derivations. This will affect the exterior derivative $d$. In
other words, if we are given a basis of 1-forms and an algebra of
derivations for the noncommutative algebra,  we may still write
$d$ as  $d=\theta^a X_a$ but it is not true in general that we can
perform a change of bases both for one-forms and derivations such
that the same exterior derivative $d$ is also equal to some
$\alpha^a
 Y_a$. Indeed, once we have performed the change
 of basis for the one-forms (which we can do, the one-forms being a
 ${\mathcal A}$-module)  we cannot rearrange the derivations in order
 that they stay derivations, apart from multiplying them by numbers or
  elements in the centre of ${\mathcal A}$. The main point of the
  paper is therefore the construction of a differential calculus for
  subalgebras of the Moyal algebra on $R^4$, $\mathcal{M}_\theta$, starting from the
  definition of a differential calculus on the Moyal algebra which is suitable to be
  reduced.

\section{Differential calculus for (noncommutative) associative
algebras} For an associative algebra a differential calculus can
always be defined algebraically, once a Lie algebra of
derivations, ${\mathcal L}$, is given (see for example
\cite{segal,marmo}). A 1-form $\alpha$ is a linear map from
${\mathcal L}$   to ${\mathcal A}$. An exterior derivative $d$ is
defined as \be d\alpha (X,Y)= \rho(X)(\alpha(Y))-
\rho(Y)(\alpha(X)) - \alpha([X,Y])\label{higherforms} \ee If
$\rho:{\mathcal L} \rightarrow Der({\mathcal A})$ is a Lie algebra
homomorphism, then $d^2=d\circ d$ is zero. Higher forms are
defined as skew-symmetric multilinear maps from ${\mathcal L}$ to
the associative algebra ${\mathcal A}$.
 Thus, to
define a differential calculus on a noncommutative algebra, we
need to choose a set of derivations, that have to be independent
and sufficient, and a representation of ${\mathcal L}$ on
${\mathcal A}$.  (A set of derivations is said to be sufficient
when  the only elements which are annihilated by all of them are
in the centre of the algebra). That is, we need ${\mathcal L}$,
$\rho$ such that \be \rho(X)\left( f\ast g\right) = \left(\rho(X)
f\right) \ast g + f\ast\left(\rho(X)  g\right), ~~~ X \in
{\mathcal L},~~ f,g\in {\mathcal A} \ee where $\ast$ is the
noncommutative product in ${\mathcal A}$. Assuming such structures
are given, the first step for the construction of a differential
calculus is the identification of zero forms with the algebra
itself \be \Omega^0=\mathcal{A}.\ee Then the exterior derivative
is implicitly defined by \be  df (X)=\rho(X) f\label{d}\ee It
automatically verifies the Leibnitz rule because $\rho(X), ~ X\in
{\mathcal L} $ are $\ast$-derivations \be d(f\ast g)(X)=
\left(\rho(X) f\right)\ast g + f\ast\left(\rho(X) g\right)\ee
moreover \be d^2=0 \label{d2} \ee because the $\ast$-derivations
$\rho(X), ~ X\in {\mathcal L}$ close a Lie algebra. The second
step consists in defining $\Omega^1$ as a left $\mathcal{A}$
module that is \be g df(X)=g\ast(\rho(X) f).\ee Analogously we can
define a right $\mathcal{A}$ module. Because of noncommutativity
they are not the same, but we can always express one in terms of
the other. Thus, we consider left modules from now on. To
construct $\Omega^2$ we use \eqn{higherforms} and \eqn{d2}. We
have \be df \diamond dg(X_\mu,X_\nu) = df(X_\mu)\ast df(X_\nu)
-df(X_\nu)\ast df(X_\mu)\ee where we have indicated with
$\diamond$ the  product of forms. Because of noncommutativity \be
df\diamond dg \ne - dg\diamond df. \ee In a similar way to
$\Omega^1$, $\Omega^2$ is defined as a left $\mathcal{A}$ module
with respect to the $\ast$ multiplication \be f dg \diamond
dh(X_\mu,X_\nu)=f\ast dg(X_\mu)\ast dh(X_\nu) - f\ast
dg(X_\nu)\ast dh(X_\mu).\ee Higher $\Omega^p$ are built along the
lines of the commutative case.

\section{A differential calculus for the Moyal algebra}
The simplest and mostly studied noncommutative algebra is the
Moyal algebra. This is a  deformation of the  algebra of functions
on $\real^{2n}$, $ (\mathcal{F}(\real^{2n}), \cdot)$ into  the
noncommutative algebra $(\mathcal{M}, \ast_\theta)$ where
$\ast_\theta$ is the Moyal product \cite{Gr46,Mo49} and $\theta$
the noncommutativity parameter. The zero-th order in $\theta$
yields back the ordinary commutative product, while the first
order is the Poisson bracket which we assume for simplicity the
canonical one.  Different (nondegenerate) Moyal products on
$\real^4$ are in principle associated with an invertible
antisymmetric matrix $\Theta_{ij}$ which, with a change of
coordinates, can be expressed in the canonical form: \be
\Theta=\left(
\begin{array}{cccc}
0& 0& -\theta_1&0\\0&0&0& -\theta_2\\ \theta_1 & 0 & 0 & 0 \\ 0&
\theta_2 &0&0
\end{array}
\right), \ee with ${\pm}\theta_i$ the eigenvalues of $\Theta$.  A
simple rescaling can then equate $\theta_1=\theta_2=\theta$.   In
this setting, the Moyal product $f\star_\theta g$ of two Schwartz
functions $f,g$ on $\real^4$ is defined by
\begin{equation}
f \ast_\theta g(u) := \int_{\reals^4}\int_{\reals^4}\,
L^\theta(u,v,w)\,f(v)g(w) \,d\mu^\theta(v)\,d\mu^\theta(w),
\label{eq:Moyal-prodint}
\end{equation}
where $u:=(q,p);\;\theta$ is a positive real parameter;
$d\mu^\theta(v):= (\pi\theta)^{-4}\,d\mu(v)$. The integral kernel
$L^\theta$ is given by
\begin{equation}
L^\theta(u,v,w):= \exp\biggl(\frac{2i}{\theta}\bigl(uJv + vJw +
wJu\bigr)\biggr), \label{eq:Moyal-prod-ker}
\end{equation}
where $J$ denotes the antisymmetric matrix: \be
J:=\left(\begin{array}{cc} 0 & \id_2 \\ -\id_2 & 0
\end{array}\right), \label{laJ} \ee with $\id_2$ the $2{\times} 2$
identity matrix.  What is properly defined as the Moyal algebra is
$\mathcal{M}_\theta:= \mathcal{M}_L({\real}_\theta^4)\cap
\mathcal{M}_R({\real}_\theta^4)$ where
$\mathcal{M}_{L}({\real}_\theta^4)$, the left multiplier algebra,
is defined as the subspace of tempered distributions that give
rise to Schwartz functions when left multiplied by Schwartz
functions;  the right multiplier
algebra~$\mathcal{M}_R({\real}_\theta^4)$ is analogously defined.
For more details we refer to the appendix in \cite{GLMV02} and
references therein. In the present article  we  shall think of
$\mathcal{M}_\theta$ as the algebra of $\ast$-polynomial functions
in $q_i,p_i,$ properly completed. Its commutative limit,
$\mathcal{F}(\real^4)$, is the commutative multiplier algebra
$\mathcal{O}_M(\real^4)$, the algebra of smooth functions of
polynomial growth on $\real^4$ in all derivatives \cite{GGISV04}.
To define a differential calculus in the constructive way
described in the previous section we need derivations. The
$\mathcal{M}_\theta$ are normal spaces of distributions, and all
their derivations are inner. Therefore we turn our attention to
groups of automorphisms of $\mathcal{M}_\theta$. A relevant one is
the inhomogeneous symplectic group  ISp(4,\real), constituted by
translations  and real symplectic transformations of
$\real^4$.\footnote{Note however that smaller Moyal algebras can
be chosen, such that the inhomogeneous symplectic algebra acts as
outer derivations on them \cite{GGISV04}. The choice of such big
algebras in the present paper is motivated by the fact that they
contain all polynomials.}
 As we
will see below in more detail, it induces  derivations both for
the commutative algebra $\mathcal{F}(\real^4)$ and the Moyal
algebra ${\cal M}_\theta$.  In facts its Lie algebra is the
maximal algebra of derivations with this property. Moreover,
although it is not minimal (the subalgebra of translations would
suffice) it generates  the whole algebra of polynomial functions,
once we represent its generators as quadratic-linear functions in
$\real^4$.

The group $Sp\left(4,\real\right)$ consists of elements $g$ for
which $g^{t}Jg =J$; this implies for the Lie algebra generators
that $M^{t}J+J M=0$, with \be \left[M_a,M_b\right]=C_{ab}^{c}M_c
\ee and  $C_{ab}^{c}$ the structure constants of the symplectic
algebra. The Lie algebra is realized in terms of vector fields on
$\real^{4}$ by: \be Y_{a}\,=\,-{M_a}^{\mu}_{\nu}
u^{\nu}\frac{\del}{\del u^{\mu}}. \ee To the symmetric matrices $
B_{a}\,=\,-J M_{a} $ it is associated a set of quadratic functions
on $\real^{4}$: \be y_{a}\,=\,\frac{1}{2}u^{t}{B_a} u. \ee They
define a realization of the symplectic algebra as a Poisson
algebra with respect to the canonical Poisson bracket $\left\{q_i,
p_j\right\}= \delta_{ij}$: \be \{y_{a},y_{b}\}=C_{ab}^c y_{c}. \ee
The inhomogeneous sector of the Lie algebra of ${\rm
ISp}\left(4,\real\right)$ is represented by linear functions.
Therefore the whole inhomogeneous symplectic algebra,
$isp\left(4,\real\right)$, may be realized as a Poisson algebra on
$\real^4$ with generators a set of quadratic-linear functions of
$q, p$.
 A possible choice for the generators is \beqa
y_1&=&{1\over 2} (q_1 q_2 +p_1 p_2) ,~~~~
y_2=\half(q_{1}p_{2}-q_{2}p_{1}),~~~~y_3={1\over
4}(q_1^2+p_1^2-q_2^2-p_2^2)\nn\\ y_4&=&{1\over
4}(q_1^2+p_1^2+q_2^2+p_2^2),~~~~y_5={1\over
4}(q_1^2+q_2^2-p_1^2-p_2^2),~~~~y_6=\half(q_{1}p_{1}+q_{2}p_{2})\nn\\
y_7&=&\half(q_{1}p_{2}+q_{2}p_{1}),~~~~y_8={1\over 2}(q_1p_1-q_2
p_2),~~~~y_9={1\over 2} (q_1 q_2 -p_1 p_2),\\ y_{10}&=&{1\over
4}(q_1^2-q_2^2-p_1^2+p_2^2),~~~~y_{11}=q_1,~~~~y_{12}=q_2\nn\\y_{13}&=&p_1,~~~~y_{14}=p_2.
\label{quadlin} \eeqa We have (latin indices now run from 1 to 14)
\be \{y_a, y_b\}=C_{ab}^c y_c,
 \label{poisson-isp}\ee
 with $C_{ab}^c$ the structure constants of the whole $isp(4,R)$. Thus,  the generators of the Lie
algebra $isp(4,\real)$, act as inner derivations in  $
\mathcal{F}(\real^4)$ with \be \rho(M_a) (f) =
Y_a(f)=\{y_a,f\}.\ee Let us notice that the vector fields $Y_a$
are the Hamiltonian vector fields associated to the functions
$y_a$; therefore to the linear functions $q_i,p_i~ i=1,2$ we
associate the vector fields $\del/\del p_i, -\del/\del q_i$
 respectively. The algebra of
quadratic-linear functions of $q, p$ is also closed with respect
to the Moyal product: using  the asymptotic development, which
becomes exact when at least one of the two elements of the product
is quadratic-linear, it is possible to show \cite{GLMV02} that the
product of two such functions is still a function of $\{y_a\}$:
\be y_a \ast y_b =f(\{y_c\}). \label{starisp}\ee Moreover the
Moyal bracket or $\ast$-commutator essentially coincides with the
Poisson bracket \eqn{poisson-isp} \be [y_a, y_b]_\ast = i\theta
C_{ab}^c y_c.\label{quantumisp}\ee Thus, the generators of the Lie
algebra $isp(4,\real)$,  act as inner derivations in  $
\mathcal{M}_\theta$ as well, with \be \rho(M_a) (f) =
[y_a,f]_\ast, ~~ f\in \mathcal{M}_\theta \ee with the Leibniz rule
trivially satisfied. Thus, $isp(4,\real)$ plays the double role of
generating $\mathcal{M}_\theta$ and furnishing $\ast$-derivations;
moreover, this is true in the commutative limit as well.
 According to the general procedure outlined in the previous section,
 once we have
derivations we can define an exterior derivative $d$ and construct
a differential calculus on  ${\cal M}_\theta$ which is certainly
not minimal, but has interesting properties. The idea we want to
pursue, which will be developed in the next section, recalls very
much the construction of a differential calculus on the algebra of
$N\times N$ matrices described by Madore in \cite{madorebook}.
There, a redundant calculus is constructed  which is what is
needed to define differential calculi for different subalgebras of
Mat(N). Here, after identifying many interesting subalgebras of
$\mathcal{M}_\theta$ we will define a differential calculus for
each of them, the main difference with the previous case being
that our algebras are realized as operators on infinite
dimensional space.

\section{A differential calculus for subalgebras}

 The algebra $ \mathcal{M}_\theta $
 has interesting subalgebras, which we indicate  generically with $\mathcal{B}$, which share
 the same commutative limit, $\mathcal{F}(\real^3)$. Therefore we
 regard them as different deformations of $\mathcal{F}(\real^3)$,
 each of them with its own $\ast$-product.
 Those subalgebras  are polynomially generated by 3-d subsets of
 the quadratic linear functions $y_\mu$ given by \eqn{quadlin}. It can be shown that they
 are in one to one correspondence with 3-d Lie algebras
 \cite{MMVZ94} which they realize both as Poisson algebras \cite{MMVZ94} and as
 $\ast$ algebras \cite{GLMV02}.
We briefly review the procedure followed in \cite{GLMV02}
 for the convenience of the reader.
Consider first the identification ${\mathcal G}^*\equiv\real^3$,
where ${\mathcal G}^*$ is the dual algebra of some three
dimensional Lie algebra. It is known that all three dimensional
algebras can be classified and a Poisson realization can be given
once the generators of a Lie algebra are identified with the
linear functions on the dual \cite{GMP93,MMVZ94}. In normal form
we have \be \{x,y\}=cw+hy,~~~~ \{y,w\}=ax,~~~~ \{w,x\}=by- h w
\label{classbr} \ee where $a,b,c,h$ are real parameters
characterizing the algebras and satisfying the condition $ah=0$.
Choosing appropriately the parameters we reproduce all the 3-d Lie
algebras.

Consider now $\real^4$ with the canonical symplectic structure
given by the Poisson brackets
$$
\{q_i,p_j\} = \delta_{ij},
$$
associated to the symplectic form
$$
\omega = dq_1\wedge dp_1 + dq_2\wedge dp_2.
$$
It is possible to find  symplectic realizations
$\pi:\real^4\rightarrow{\mathcal G}^*\equiv\real^3$, for
${\mathcal G}^*$ dual to any 3-dimensional Lie algebra, ${\mathcal
G}$. We express $\pi$ through the change of variables $\pi^*$ that
pulls smooth functions on $\real^3$ back to smooth functions on
$\real^4$.
 All that one has to do is to find three
independent functions $f_1,f_2,f_3$ on $\real^4$ whose
corresponding canonical brackets have the required
form~\eqn{classbr}. The Poisson map $\pi$ is {\it not\/} required
to be onto, nor a submersion, that is to say, to arise from a
regular foliation of $\real^3$.

Several $\pi$-maps were constructed in~\cite{MMVZ94}, under the
name of (generalized) classical Jordan--Schwinger maps.  Although
many realizations are possible,  it turns out that it is always
possible to find a realization for every $\mathcal{G}$ in terms of
quadratic-linear functions on $\real^4$, namely as a Poisson
subalgebra, $\mathcal{B}$,  of the  algebra $isp(4)$ given by
\eqn{quadlin}.

Less obvious is that the subalgebras are also closed under the
induced  $\ast$-product. We have indeed   (now Latin indices run
from 1 to 3) : \beqa y_i\ast y_j&=& f(\{y_i\}) \nn \\
 \left[y_i,y_j\right]_\ast &=& i \theta \{y_i, y_j\}.
 \eeqa
That is, for each set of generators we observe that:
\begin{itemize}
 \item they generate polynomially a noncommutative subalgebra of $\mathcal{M}_\theta$, say $\mathcal{B}_{\mathcal{G}}$;
 \item they close the Lie algebra $\mathcal{G}$, which is a subalgebra of $isp(4,\real)$,  both
 with respect to the Moyal bracket  and to the Poisson bracket;
 \item this implies that $\mathcal{G}$ acts on $\mathcal{B}_{\mathcal{G}}$ in terms of  inner
 $\ast$-derivations
 \be
 \rho(M_i) f = [y_i, f]_\ast,~~~ M_i\in \mathcal{G},~~f\in \mathcal{B}.
\label{astder}
 \ee
 \end{itemize}
A detailed account of all the subalgebras of \eqn{quadlin} and
related star products is contained in \cite{GLMV02}. We shall only
recall that there are essentially two families of subalgebras.
Those that we call of type A in the cited article, which are
defined by the property of being the commutant of a certain
function in the list \eqn{quadlin}, which can be identified as the
Casimir function, it corresponding exactly to the Casimir of the
associated Lie algebra. To this class belongs the example we study
below. To the  other broad class belong the so called  type B
algebras, that is, algebras defined through a Casimir 1-form which
is not exact. Among them, an interesting case is the k-Minkowski
algebra which describes a deformed 2+1 Minkowski space. The
differential calculus that we construct is essentially different
for the two cases. We will see that for type A algebras it is
possible to find a frame of 1-forms which behave as in the
commutative case, whereas for type B algebras this is not
possible.

As a guiding example we shall refer, when needed, to the type A
subalgebra generated by $y_1, y_2, y_3$ as in eq. \eqn{quadlin}.
As a Poisson algebra this is easily seen to be isomorphic to
$su(2)$. More precisely, it is the commutant of $y_4$. The induced
star product is in that case \be y_j \ast_{su(2)} f(y_i) = \{y_j
-{i\theta\over 2}\epsilon_{jlm} y_l \del_m - {\theta^2\over 8}[(1
+ y_k \del_k)
\partial_j - {1\over 2} y_j \,\del_k \del_k]\} f(y_i).
\label{starsu2} \ee The nonlocality of this product is evident
once we observe that $y_i\ast y_i= y_i^2 -{1\over 8} \theta^2$ (no
sum over repeated indexes). The algebra generators may be
represented in terms of creating and annihilating operators acting
on  the usual Hilbert space of the two dimensional harmonic
oscillator, with basis the cartesian kets $|n_1 n_2>$. The sum
$n_1+ n_2$ is constant, it being the eigenvalue of $y_4$, which
commutes with the whole algebra and represents the Hamiltonian of
the system of oscillators. Therefore, changing basis to the
$\{$Hamiltonian + angular momentum$\}$ basis, it is possible to
see that for each value of the angular momentum  there is a
representation of su(2). The noncommutative algebra of functions
of $R^3$ therefore reduces to a set of finite dimensional
algebras, receptacles for representations of su(2).  Each reduced
block is the algebra of a fuzzy sphere \cite{Madorefuzzysphere} in
the oscillator representation. Therefore the three dimensional
space is ``foliated" as a set of fuzzy spheres of increasing
radius. We can give a geometric interpretation of the new star
product. Note that, with the exception of the zero orbit, the
orbits of the Hamiltonian system associated to $y_4$  are circles.
Functions of $(y_1,y_2,y_3)$ correspond here to functions of
$(q_1, q_2, p_1, p_2)$ that remain invariant on those orbits. We
are thus identifying $R^3$ to the foliation of $R^4$ by those
trajectories. The orbits  rest on spheres in $R^4$. One circle and
only one passes through each point different from 0. The
corresponding maps $S^3 \rightarrow S^2$ are Hopf fibrations.

 A differential calculus on each
$\mathcal{B}_\mathcal{G}$ is straightforward to define along the
same lines of the previous section. This is the natural reduction
of the differential calculus on $\mathcal{M}_\theta$ to the
subalgebras $\mathcal{B}_{\mathcal{G}}$. In particular the
exterior derivative may be  defined as \be d_{\mathcal{B}} f
(M_i)= [y_i, f]_\ast \label{ddef}\ee with $f, y_i \in
\mathcal{B}_{\mathcal{G}}$, $M_i \in \mathcal{G}$. If $\phi
:\mathcal{B}_{\mathcal{G}}\rightarrow \mathcal{M}_\theta$ is the
embedding in the Moyal algebra we have \be \phi (d_{\mathcal{B}}
f) =d_{\mathcal{M}} \phi (f) \label{proj}\ee that is, the
differential calculus we have defined on the Moyal algebra induces
a differential calculus on subalgebras. The condition for that to
be possible in the noncommutative case is that the derivations we
have chosen for $\mathcal{M}_\theta$ be `adapted' to those of
$\mathcal{B}$. Had we chosen as an algebra of derivations just the
translations, which is what gives the minimal differential
calculus on the Moyal algebra, Eq. \eqn{proj} wouldn't have been
true. This justifies a posteriori our choice of such a big
calculus for $\mathcal{M}_\theta$.
\subsection{Frame of 1-forms}
For each fixed subalgebra $\mathcal{B}_{\mathcal{G}}$ the set of
$\{dy_i\}$ certainly constitutes a system of generators of
$\Omega^1(\mathcal{B}_{\mathcal{G}})$ but it is not the most
convenient one. Because of noncommutativity we have indeed $f(y_i)
dy_j \ne dy_j f(y_i)$. A better system of generators for
$\Omega^1$  would be 1-forms which are dual to the derivations
$\rho(M_i)$.   Finding a frame of 1-forms for the subalgebras
$\mathcal{B}_\mathcal{G}$ is not always possible. In facts, there
is no solution in all the cases where the algebra has neither a
centre nor a unity (type B algebras of \cite{GLMV02}),  but we
will show that a solution exists for the subalgebra considered
above.

Once we have found the frame, the construction of the differential
calculus follows very closely the construction of Madore
\cite{madorebook} for finite-dimensional (matrix) algebras,
although ours is not finite-dimensional. Differential calculi
constructed in this way depend on the algebra of derivations one
has chosen.
 We will show at the end of the section  how it can be made independent
 on derivations and formulated in terms of a one-form which recalls the Dirac operator
 of Connes differential  calculus \cite{connes}.

For a generic  subalgebra $\mathcal{B}_\mathcal{G}$ the system to
be solved is
 \be
(\alpha^i)(M_j)\in Z(\mathcal{B}_\mathcal{G}) \label{frameeq4} \ee
where $i=1,..3$, $M_i$ are the generators of the Lie algebra of
$\ast$-derivations of $\mathcal{B}_\mathcal{G}$ and
$Z(\mathcal{B})$ is the centre of $\mathcal{B}$. Here we have
slightly modified the definition of dual frame, because our
algebras have no unity. If the centre is trivial \eqn{frameeq4}
has no solutions. Therefore the problem is meaningful only for
type A subalgebras.

Let us consider the algebra $\mathcal{B}_{su(2)}$. To the centre
belong all functions of $y_1^2+
y_2^2+y_3^2=[(q_1^2+q_2^2+p_1^2+p_2^2)/4]^2$. Since a generic
one-form may be written as  $\sum_a f_a dg_a$ it can be easily
seen that there are no solutions which can be expanded in the
basis $\{dy_i\}$. Therefore
 we write the dual forms as
\be \alpha^i=f^i_1 dq_1 + f^i_2 dq_2+ f^i_ 3 dp_1+f^i_4 dp_2 \ee\
where $f^i$ are functions in $R^4$ and the 1-forms $dq_i, dp_i$
are defined by \eqn{d}. By means of Eq. \eqn{ddef} equation
\eqn{frameeq4} becomes then \be f^i \ast {i\theta\over 2} A = z
\delta^{ij} e_j\ee where $f^i$ is a row vector, $A$ is the  $3
\times 4$ matrix \be \left(
\begin{array}{ccc}
  -p_{2} & q_{2} & -p_{1} \\
  -p_{1} & -q_{1} & p_{2} \\
  q_{2} & p_{2} & q_{1} \\
  q_{1} & -p_{1} & -q_{2}
\end{array} \right)
 \ee
$z$ is an element  in the centre and $e_j$ is the row vector
$(0,..,1_j,0,..)$. Since the algebra $\mathcal{B}_{su(2)}$ has a
centre, solutions to \eqn{frameeq4} are in principle defined up to
one-forms in the kernel of $\{M_1,M_2,M_3\}$ (`Casimir
one-forms'). Therefore, to solve the problem we enlarge the
algebra of derivations introducing the one associated to the
generator $M_4$, which commutes with all the others. This is
represented by the quadratic function $y_4$ in the list
\eqn{quadlin}, which commutes with all the elements of our
algebra. We look then for a one-form, $\alpha^4$, dual to this
auxiliary
 derivation. If existing, it will be a Casimir one-form. The
system to be solved becomes  now \be f^\mu \ast {i\theta\over 2} A
= z \delta^{\mu\nu} e_\nu\ee where, with an obvious extension of
the notation,  the index $\mu$ runs from 1 to 4 and $A$ is the
square matrix \be \left(
\begin{array}{cccc}
  -p_{2} & q_{2} & -p_{1}& -p_1 \\
  -p_{1} & -q_{1} & p_{2}&-p_2 \\
  q_{2} & p_{2} & q_{1}&q_1 \\
  q_{1} & -p_{1} & -q_{2}&q_2
\end{array} \right)
\label{Lambda} \ee We need therefore a right $\ast$-inverse, that
is a matrix $B$ such that $A\ast B = z$.  In general we are not
guaranteed that a matrix with noncommuting entries have an inverse
in the sense specified above; in this case it exists (indeed, it
is possible to define a $\ast$-determinant, which is non-zero and
central). Notice that in the commutative limit $\det A=0$, that is
the matrix $A$ is degenerate. The solution for the frame of
one-forms is finally
 \beqa \alpha^1&=&{C} \left[i(p_2 dq_1+p_1 dq_2 -q_2 dp_1
  -q_1 dp_2) -{2 \over \theta} y_1\beta\right]\nn\\
\alpha^2&=&{C } \left[i(q_1 dq_2 -q_2 dq_1 -p_2 dp_1 +
  p_1 dp_2) -{2 \over \theta} y_2\beta \right]\nn\\
\alpha^3&=&{C } \left[i(p_1 dq_1 -p_2 dq_2 -q_1 dp_1
  +q_2 dp_2) -{2 \over \theta} y_3\beta\right]\\
  \alpha^4&=&C \left[{2 \over \theta} y_4 \beta\right]
   \eeqa where $\beta=
(q_1 dq_1 + q_2 d q_2 + p_1 dp_1 + p_2 d p_2)$,
   $\theta$ is the noncommutativity parameter and $C$ is  a normalization
   constant. The one-form $\beta$ is in the kernel of the algebra of derivations
   generated by $M_1, M_2, M_3$ (notice however that it is not
  equal to $ d(q_1^2 + q_2^2 +  p_1^2 + p_2^2)/2$), therefore it is ineffective
  as long as we are concerned with the subalgebra generated by $\{y_1,y_2,y_3\}$. The differential
  calculus which we have induced on $ \mathcal{B}_{su(2)}$ is three-dimensional and generated
  by the frame of one-forms $\alpha_1,\alpha_2,\alpha_3$. Notice that, up to the
one-form $\beta$,  these are exactly the dual 1-forms of left
invariant vector fields on the group manifold $SU(2)$ when
immersed in $R^4$. In facts, from \eqn{astder},  in
 the commutative limit the three derivations go into the vector
 fields
 \beqa
  Y_1&=& D \left (p_2{\del\over \del q_1} + p_1{\del\over \del q_2} - q_2{\del\over \del
 p_1}- q_1{\del\over \del p_2} \right ) \nn\\
 Y_2&=& D \left (- q_2{\del\over \del q_1} + q_1{\del\over \del
 q_2}-  p_2{\del\over \del p_1} + p_1{\del\over \del p_2}  \right ) \nn\\
 Y_3&=& D \left (p_1{\del\over \del q_1} - p_2{\del\over \del
 q_2}- q_1{\del\over \del p_1} + q_2{\del\over \del p_2} \right ) \label{leftinv}
 \eeqa
 which can be recognized  to be a basis of left invariant vector
 fields on the 3-sphere. These are independent if we only allow
 numerical coefficients, but not as a module. In the
 noncommutative case they are independent because there is no
 module structure. Therefore, recalling the geometric interpretation we have given of the
 representation space of
 the $ \mathcal{B}_{su(2)}$ algebra as a foliation into fuzzy spheres,
 we recover the known result that the tangent space to noncommutative
 2-spheres is three dimensional and not two dimensional.

   As
anticipated in the beginning of the section, the existence of a
frame simplifies very much the construction of the differential
calculus and makes it possible to model it on the existing
differential calculi for finite-dimensional matrix algebras, thus
allowing to recover many of the properties we have in that case.
We will enumerate some of them. Because of their definition
\eqn{frameeq4} fundamental forms verify \be f\alpha^i=\alpha^i
f\label{comm}. \ee Then $\Omega^1({\mathcal B})$ is  a free module
of rank 3. Moreover $\alpha^i\diamond \alpha^j=-\alpha^j\diamond
\alpha^i$ which implies that forms of degree higher than 3 vanish.

From the same equation \eqn{frameeq4}  we derive the Lie
derivative: \be 0={\mathcal L}_{Y_i} <Y_j,\alpha^k>= <({\mathcal
L}_{Y_i} Y_j), \alpha^k> +  <Y_j,{\mathcal L}_{Y_i} \alpha^k>. \ee
The Lie derivative of a derivation being just the Lie bracket we
have then \be {\mathcal L}_{Y_i} \alpha^k= \alpha^l \epsilon_{l i
}^k . \label{Lieder1-forms} \ee  From the definition of exterior
derivative \eqn{d} we find an important property of fundamental
forms: \be d\alpha^i= {1\over 2} \epsilon^i_{jk}\alpha^j\diamond
\alpha^k \ee which is the Maurer Cartan equation.
  The fundamental one-forms being graded-commutative we can construct the external algebra
 $\Lambda^*$, so that
$\Omega^\ast({\mathcal B})={\mathcal B}\otimes \Lambda^\ast$.

 From
the fundamental forms $\alpha_i$ we can construct a 1-form in
$\Omega^1({\mathcal B})$ \be \alpha=-y_i \alpha^i \ee in terms of
which we can reexpress the exterior derivative $df$
 as
\be df = -[\alpha,f]. \ee Here there is no explicit reference to
derivations. The 1-form $\alpha$ generates $\Omega^1({\mathcal
B}_{su(2)})$ as a bimodule.

The construction of $\Omega^1({\mathcal B}_{su(2)})$ that we have
presented in this section may be easily repeated for  the
subalgebra ${\mathcal B}_{su(1,1)}$. The other type A algebras may
be obtained as contractions of either ${\mathcal B}_{su(2)}$, or
${\mathcal B}_{su(1,1)}$, therefore it should be possible to
generate for them a frame of 1-forms through a contraction
procedure.

\section{Concluding remarks}
In this paper we have addressed the problem of defining a
differential calculus for noncommutative algebras possessing a
sufficient number of $\ast$-derivations.  To this purpose we have
reviewed a procedure due to Segal to define a differential
calculus for the algebra of operators of Quantum Mechanics, where
the main ingredient was the existence of a Lie algebra of
derivations. Inspired by an existing construction for matrix
algebras due to Madore, we have found for a relevant case a frame
of 1-forms and discussed the commutative limit.

 \end{document}